# Design of the Second-Generation ARIANNA Ultra-High-Energy Neutrino Detector Systems

Stuart A. Kleinfelder, *Senior Member, IEEE*, for the ARIANNA Collaboration[1]

*Abstract*— We report on the development of the seven station ARIANNA Hexagonal Radio Array neutrino detector systems in Antarctica. The primary goal of the ARIANNA project is to observe ultra-high energy (>100 PeV) cosmogenic neutrino signatures using a large array of autonomous stations each dispersed 1 km apart on the surface of the Ross Ice Shelf. Sensing radio emissions of 100 MHz to 1 GHz, each station in the array contains RF antennas, amplifiers, a 2 G-sample/s signal acquisition and trigger circuit I.C. (the "SST"), an embedded CPU, 32 GB of solid-state data storage, a 20 Ah $LiFePO_4$ battery with associated battery management unit, Iridium short-burst messaging satellite and long-distance WiFi communications. The new SST chip is completely synchronous, contains 4 channels of 256 samples per channel, obtains 6 orders of magnitude sample rate range up to 2 GHz acquisition speeds. It achieves 1.5 GHz bandwidth, 12 bits RMS of dynamic range, ~1 mV RMS trigger sensitivity at >600 MHz trigger bandwidth and ps-level timing accuracy. Power is provided by the sun and $LiFePO_4$ storage batteries, and the second-generation stations consume an average of 4W of power. The station's trigger capabilities reduce the trigger rates to a few milli-Hertz with ≤4-sigma thresholds while retaining high efficiency for neutrino signals.

## I. INTRODUCTION

THE ARIANNA project (Antarctic Ross Ice-shelf ANtenna Neutrino Array) is a surface array of radio receivers planned to span of order 1,000 $km^2$ of the Ross Ice Shelf in Antarctica, viewing ~0.5 Teratons of ice [1]. The project will detect radio waves originating from high energy neutrino interactions with atoms in the ice via the Askaryan Effect.

An almost complete system redesign has been made and deployed during the 2014 and 2015 campaigns. This has included updated power and communications tower systems (Fig. 1, left), the design of a new fast sampling chip, the development of a new single-board system to replace the previous motherboard/daughter-card system (Fig 1, right), an updated amplifier design, and new battery systems. The resulting hardware has improved electrical and physical robustness, better features and performance, uses substantially less power, is less costly, and is easier to calibrate. It maintains full "drop-in" compatibility with the previous-generation HRA systems [2].

## II. THE SST, A NEW 2 GS/S WAVEFORM RECORDER I.C.

A new signal sampling and triggering integrated circuit has been designed and fabricated [3, 4]. Containing 4 channels of 256 samples per channel, the "SST" Synchronous Sampling plus Triggering I.C. incorporates substantially the same functionality as the ATWD system in [2], but in a greatly simplified, easier to use, higher-performance and lower-power form. The sampling is completely synchronous, using no PLL or any delay-based timing, and is simply driven by an external LVDS clock for high timing uniformity and stability. Optimized design and packaging yielded a nearly-flat analog bandwidth to ~1.3 GHz using a standard 50-Ohm signal source and a -3 dB bandwidth of ~1.5 GHz. The use of an inexpensive 0.25 μm CMOS process allows a large input voltage range of 1.9V on a 2.5V supply, reaching 12 bits RMS dynamic range. The SST requires no programming, and only 3 active control signals are required to operate it: Reset, Run/Stop, and Read-Clock. The power consumption of the chip depends on the clock rate, the duty cycle of acquisition vs. digitization, and the bias on the comparators. When operating at the HRA's normal 2 G-samples/s acquisition speed and with typical trigger settings, power consumption is about 32 mW per channel.

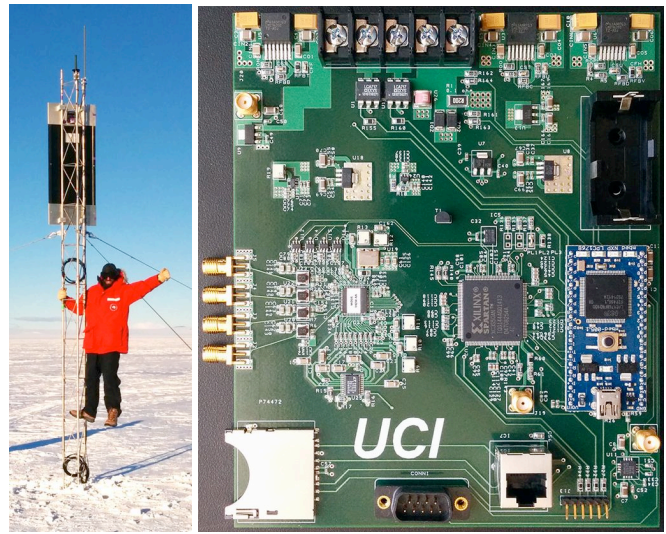

Fig. 1. Left: A "new" ARIANNA power tower including a single 100W solar panel, Iridium satellite communications and long-distance wireless antennas. Right: The ARIANNA single-board data acquisition system, including one four-channel SST chip (small square chip with a white label, left of center), improved power management and on-board temperature monitoring. Average power consumption of this board is 1.7W.

[1] This work was supported by funding from the Office of Polar Programs and Physics Division of the U.S. National Science Foundation, grant awards ANT-08339133, NSF-0970175, and NSF-1126672, by NASA via grant 14-
S. A. Kleinfelder is with the Department of Electrical Engineering and Computer Science at the University of California, Irvine, CA, 92697, U.S.A., 949-824-9430, stuartk@uci.edu.

An example waveform acquired by the SST and the system board seen in Fig. 1 is shown in Fig. 2. This is a 100 MHz sine wave plotted after voltage and timing corrections. Timing calibration was obtained by simulated annealing [4].

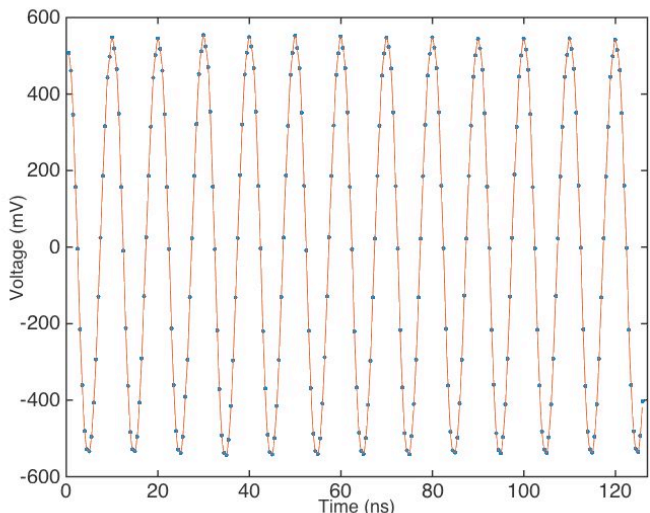

Fig. 2: 100 MHz sine wave acquired at 2 GHz after voltage and timing corrections.

The SST includes a per-channel dual-threshold windowed coincidence trigger (Fig. 3) that operates with 1 mV RMS resolution and >600 MHz equivalent input bandwidth (e.g., it is sensitive to small-signal pulses down to 500 ps FWHM or better, with 0% to 100% triggers spanning less than 4 mV in pulse height differences). An AND or an OR can be formed between comparators per channel over a window of ~3.5 ns or greater to form a bipolar trigger. For example, if set to 5 ns, a coincidence equivalent to a bipolar signal of greater than 100 MHz can pass this first-level trigger. Output pins are available for each individual trigger comparator for easy calibration and rate monitoring or else, during typical operation, the AND of each channel's two comparators can be output in differential form. Trigger outputs can also be set to reduced output voltages, e.g. positive ECL (0-0.8V), for low cross-talk.

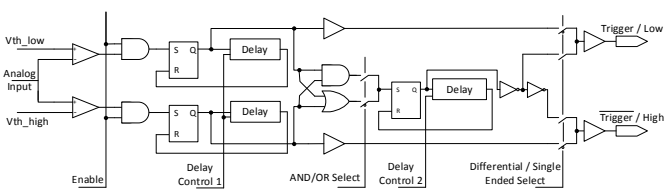

Fig. 3: Trigger logic including enable, pulse-stretching latches and delay lines, AND/OR selection and pulse-stretching, and selection of dual single-ended or differential outputs.

Fig. 4 shows an example trigger performed by the SST chip. An 8 mV input pulse with 500 ps full-width at half-maximum is seen in the upper trace (as recorded by a 1 GHz bandwidth, 5 G-samples/s oscilloscope). The lower trace shows the SST's trigger output, which was arbitrarily set to 17 ns output width. The width can be tuned to combine channels in a second-level firmware trigger, e.g. via majority logic, to reduce trigger rates due to thermal noise. In ARIANNA, the width is typically set to 30 ns, which is a little longer than the antenna baseline.

Fig. 5 shows the digitized readout of an Askaryan-effect neutrino template produced by an arbitrary waveform generator and transmitted to two SST channels, one delayed by 6.1 ns. After corrections for voltage and timing pedestals, a cross correlation is computed, as seen in Fig. 6. The SST system boards resolve the time difference between these templates, tested over a range of delays, to better than 6 ps RMS in all measured cases.

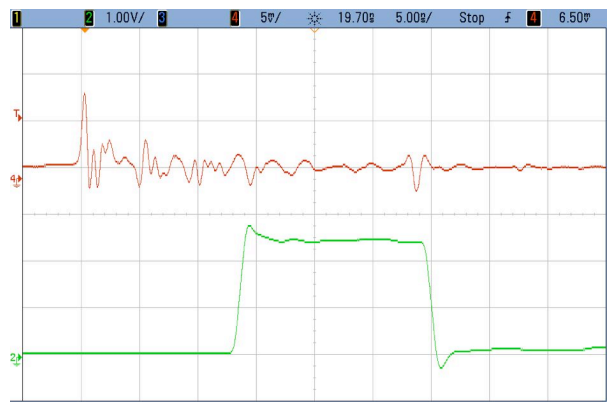

Fig. 4: An 8mV high, 500 ps wide (FWHM) input pulse (top trace) and the SST's trigger output response (bottom trace) while set to 2.5V single-ended output and ~17 ns output width.

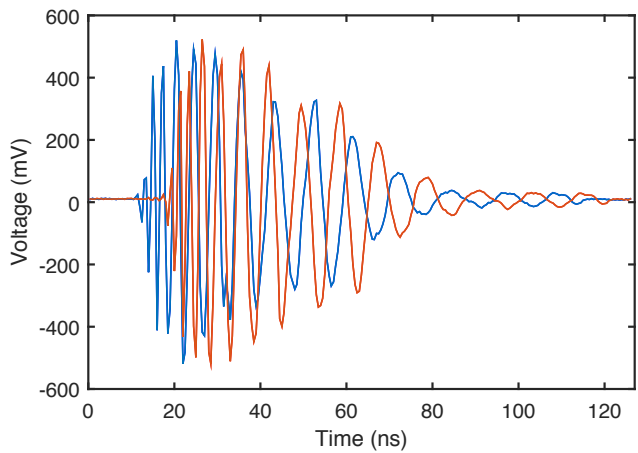

Fig. 5: Askaryan-effect template acquired by the system board seen in Fig. 1 by two channels, one delayed by 6.1 ns relative to the other.

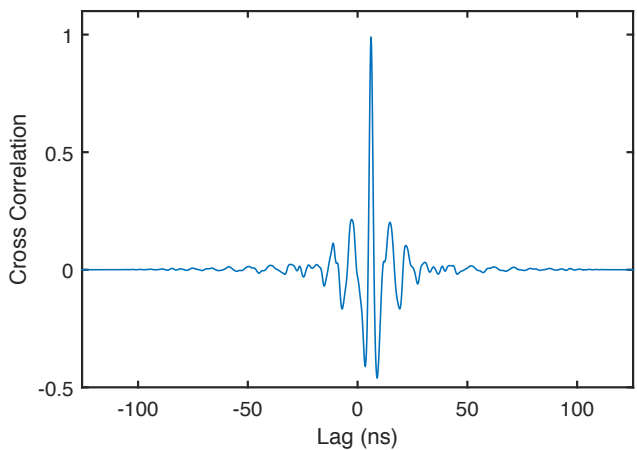

Fig. 6: Cross-correlation lag of the two waveforms seen in Fig. 3. The measured lag is 6.1 ns. Analysis of 1,000 such events, each arriving randomly in time across the 256 samples, and after time and voltage calibrations, yields under 6 ps RMS of timing imprecision.

## III. HRA System Enhancements

### A. Updated power and communications towers.

Experimental trials of inexpensive wind turbines for the supply of power during the Austral winter had demonstrated mechanical unreliability during fierce Antarctic storms and have been abandoned. Due to this, the tower configuration (Fig. 1, above) was changed to standard straight segments without a pyramidal support portion for the wind turbines, allowing the solar panels to be mounted higher on the towers. In addition, the Iridium and long-distance WiFi antennas could then be mounted at the top of the towers instead of on a separate mast. The new towers were deployed with the station electronics at their feet, at the center of the station's antenna array, and steel tower tie-down cables were replaced with R.F.-transparent Aramid cabling.

### B. Battery challenges and changes.

For wind power to be viable, large capacity batteries are necessary to bridge between windy periods. Earlier station implementations were thus equipped with 112 or 224 Ah of battery capacity (nominal rating at room temperature). But with the deletion of wind power, only a small battery is necessary to bridge days of waxing or waning sun or to supplement power during dark, cloudy days. ARIANNA thus had the opportunity to redesign the battery systems to in-house specifications.

Tests comparing several battery technologies were made, including lithium polymer batteries and two varieties of $LiFePO_4$ battery cells. Cells produced by A123 Systems were selected as having the best performance at low temperatures among those tested, retaining 60% of their nominal capacity at -30 °C while discharging at high rates (5 times the expected current demand) and also surviving tests at -60 °C.

Lithium batteries require sophisticated battery management, with especially critical concerns at cold temperatures. A Texas Instruments bq40z60-based battery management unit (BMU) was selected. This is highly programmable, including the ability to define safe high and low voltage cutoffs (when "full" and "empty") and to set a maximum charging current. For the sake of longevity at low temperatures, the maximum charging current was set to a conservative 0.08 C, meaning a current that is 0.08 times the battery pack's nominal capacity in Ah, or 1.6A maximum in this case. This is a modest percentage of the power generating capacity of the 100 W solar panel used to power the station and charge the battery (~7.5A at 13.2V while charging). In another important point, the BMU's off-current was measured to be 1.7 µA, leading to nearly negligible discharge when idle during the Austral winter.

The resulting battery was configured as 20 Ah in a 4-series, 8-parallel cell arrangement (total of ~256 Wh at room temperature). This battery, including its battery management unit, was compact enough to fit entirely within ARIANNA's existing system housing (see [2]) along side the new, smaller, single system board. The battery thus benefits from the moderate warmth generated by the operating electronics. The housing is R.F.-tight and thus also contains any noise from the BMU. At this point, *all* electronics save for the antennas and solar panels are enclosed in the system boxes.

Fig. 7 shows the operation of the resulting combination of battery, BMU and the ARIANNA system electronics when discharged at -30 °C (after having been charged at the same temperature). This is a worst-case test since all electronics in the station box were turned on at once, including both communications modules, amplifiers and data acquisition, etc. About 22 hours of operation was achieved. Using more normal operation cycles, about twice that duration is expected. This is sufficient to provide significant supplementary power during periods of waxing and waning sun, etc., to the degree that nearly 100% operation time is expected while any sun is available.

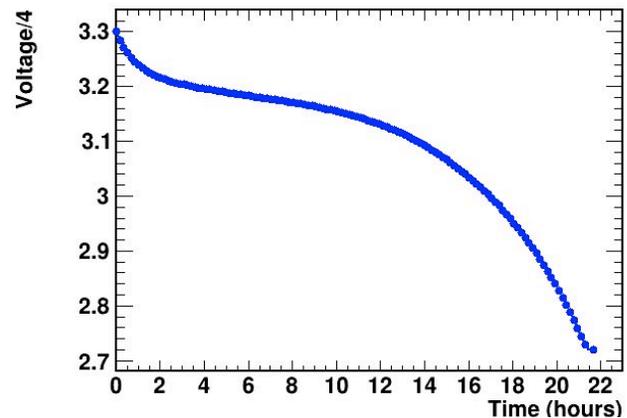

Fig. 7: Battery discharge voltage profile at -30 °C in a worst-case station operation profile (everything powered on at once).

### C. Updated amplification.

An updated amplifier has been designed (Fig. 8) and deployed in 2014. It features enhanced stability, flatter frequency response and more symmetrical gain. It also eliminates the need for external band-pass and limiting components, with at most a single attenuator needed to match its output to the system board's input range.

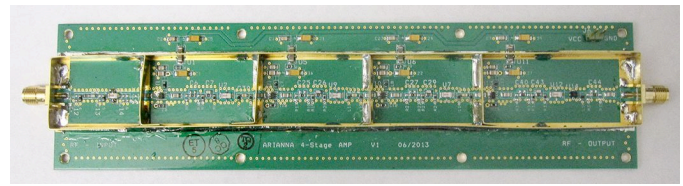

Fig. 8: Improved ARIANNA amplifier with cover shield removed.

### D. Next-generation system board.

A next-generation single-board data acquisition system board has been created and used for the completion of the HRA in 2014. The new system, seen in Fig. 1, includes one on-board 4-channel SST chip (see Section II below) in place of four daughter-cards. Power management has been improved, achieving higher input voltage tolerance (42V) along with the incorporation of on-board static protection. The new system board, and the SST in particular, has resulted in dramatic power reduction, from ~5.8W for the original HRA data acquisition system to ~1.7W for that seen in Fig. 1. Given the lower power consumption, the sensitivity and accuracy of the system board's voltage and current measurements was also enhanced. Finally, the board includes digital ambient temperature monitoring that is calibrated down to -55 °C.

Fig. 9 shows the results of an R.F. pulse generated by a Pockels cell driver bounced off of the bottom of the Ross Ice Shelf at a nearly direct down-and-up angle, and received by two parallel channels of an SST-equipped HRA station.

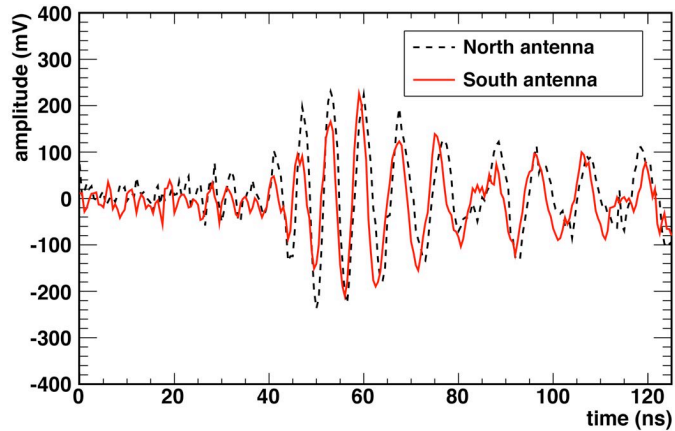

Fig. 9: Example RF pulse bounced off of the bottom of the Ross Ice Shelf and digitized by an SST-equipped HRA station.

### E. Updated system software and firmware.

The system firmware has also been significantly changed. In this version, all triggering and waveform digitization is managed autonomously by the board's FPGA. The FPGA forms second-level triggering (i.e., merging multiple channel's triggers), stops the SST, digitizes its data, and stores it in the FPGA's own block memory. Once digitization is completed, it delivers a flag to the system's microcontroller, indicating that an event has been taken and is available. This relieves the system's microcontroller from close, time-critical interaction, and permits higher levels of parallelism and lower dead-time.

The system software has also been enhanced with many improvements, including enhanced power management, a software trigger and various alternate modes of operation. For example, the two communications modes (Iridium satellite short-burst messaging and long-distance WiFi) are now individually prioritized and powered, including fall-back selection of alternate operation choices.

Trigger rates have been lowered to the milli-Hz regime by tuning thresholds and relying on the hardware coincidence features of the SST and its supervisory FPGA. This is now augmented by an optional software function which performs an FFT of each event and filters out events containing any narrow frequency spike noted to be associated with storms or airplane activity. Thermal noise and neutrino-like events have energy in a relatively broader frequency range. The deeper sample depth of the SST was useful in making this discrimination practical.

A new mode of data handling operation is also included, colloquially known as "get-one-send-one." In this mode, every time an event passes the hardware/firmware trigger and optionally the software trigger, it is immediately sent (e.g., via Iridium), prior to storage, after which data acquisition resumes. This contrasts with heretofore normal operation, in which many events are stored to flash memory first and only later retrieved for bulk transmission during fixed communications windows. It is anticipated that this new mode of operation may become the default in a very low event rate environment.

## IV. SUMMARY AND CONCLUSIONS

ARIANNA's 7 station "Hexagonal Radio Array" pilot has been completed. Significant evolution of the hardware, firmware and software has taken place. A 20 Ah LiFePO$_4$ battery pack with a highly-programmable battery management unit has been incorporated inside the system electronics box. A new single-board data acquisition system has been developed, including improved power management. The board includes the new SST analog waveform recording and triggering chip. Containing 4 channels of 256 samples per channel, the chip can record waveforms with 1.5 GHz bandwidth and 12 bits of dynamic range. Each channel contains dual-threshold discrimination able to act on signals as small as 8 mV and 500 ps in width. The SST's trigger electronics combines bipolar thresholds over an adjustable window (e.g., 5 ns for 100 MHz and above bipolar waveforms), allowing thresholds to be lowered while maintaining manageable trigger rates. An on-board FPGA then combines per-channel triggers in a programmable majority logic scheme for even lower rates. The new SST system board uses about 1.7 W, and the complete system including amplification and communications consumes about 4 W of power on average. Finally, a software trigger has been developed that examines the frequency-domain energy content of events to further reduce triggers rates.

## V. ACKNOWLEDGEMENTS

The ARIANNA Collaboration particularly thanks Anirban Samanta for the design of the SST system board and FPGA programming seen in Fig. 1, Tarun Prakash and Edwin Chiem for the design and testing of the SST chip as discussed in Section II, Chris Persichilli for the battery measurements as seen in Fig. 6, and Thorsten Stezelberger for the amplifier design seen in Fig. 7.